\def\gapp{\mathrel{\mathop{>}\limits_{{}^\sim}}}
\def\simt0{\mathrel{\mathop{\sim}\limits_{{}^{\tau\to 0}}}}
\documentstyle[12pt]{article}

\newmathalphabet{\bfmath}
\newmathalphabet{\rmmath}
\newmathalphabet{\itmath}
\newcommand{\be}{\begin{equation}}
\newcommand{\ee}{\end{equation}}
\newcommand{\ba}{\begin{eqnarray}}
\newcommand{\ea}{\end{eqnarray}}
\addtoversion{normal}{\bfmath}{cmr}{bx}{n}
\addtoversion{normal}{\rmmath}{cmr}{m}{n}
\addtoversion{normal}{\itmath}{cmr}{m}{it}

\begin{document}

\title{ About the decay of the $\Lambda$ in a nucleus }
\author{{W.M. Alberico}, {M.B. Barbaro} and {A. Molinari}
\\
\em Dipartimento di Fisica Teorica dell'Universit\`a di Torino and \\
\em Istituto Nazionale di Fisica Nucleare, Sezione di Torino, \\
\em via P.Giuria 1, I-10125 Torino, Italy}

\date{}

\maketitle

\begin{abstract}
A few aspects of the $\Lambda$ mesonic and non-mesonic decay in nuclei
are shortly addressed.
\end{abstract}

\section{The mesonic decay of the $\Lambda$}

In the following we shortly deal with the decay of the $\Lambda$
inside a nucleus.

In free space the decay of the $\Lambda$ into a pion and
a nucleon is characterized by a long lifetime ($\tau_\Lambda=2.63\times
10^{-10}$ s) because it stems from the weak interaction and its occurrence goes
together with a change of strangeness.
Yet in a nucleus, at variance with the neutron, the $\Lambda$ remains unstable.

Indeed in the former case the neutron-proton mass difference of 1.26
MeV should be compared with about 8 MeV of binding energy: hence the
general stability of the neutron
inside a nucleus and, as a consequence, of the nucleus itself.
On the contrary in the case of the $\Lambda$ one has

\be
m_\Lambda-(m_N+m_\pi) \sim 40~MeV
\ee
which is substantially larger than the $\Lambda$ binding energy in a nucleus
\cite{Mol95}:
hence the instability of an hypernucleus.

However one would like to know how nuclear structure affects $\tau_\Lambda$.
If one sticks to a pure mesonic decay, namely

\be
\label{eq:charge}
\Lambda \rightarrow p+\pi^-
\ee
or
\be
\label{eq:neutral}
\Lambda \rightarrow n+\pi^0 \;,
\ee
then it is easily recognized that when the $\Lambda$ is at rest the energy
$E_N$
and the momentum $p_N$ of the emitted nucleon are given by

\be
E_N = \frac{p_N^2}{2 m_N} \simeq \frac{(m_\Lambda-m_N)^2-m_\pi^2}{2 m_\Lambda}
\sim 5~MeV
\ee
and
\be
p_N = \sqrt{2 m_N E_N} \simeq 100~MeV/c
\ee
respectively.
Hence, conceiving the nucleus as a Fermi liquid with a Fermi momentum in the
range $220\leq k_F\leq 270 \; MeV/c$, it turns out that the $\Lambda$ decay is
Pauli blocked, which is indeed what happens already for quite low mass number,
starting, say, from $A\simeq$ 15 or even less.

A detailed calculation of the pionic $\Lambda$ decay in a nucleus has been
recently performed by Oset and Nieves \cite{Ose95}. Their findings are
displayed
in Fig.1.

\begin{figure}
\vspace{15cm}
\caption{ The mesonic width $\Gamma_\pi$ versus the free one as calculated
by Oset and Nieves [2].}
\end{figure}

The following is worth noticing in their results:
\begin{itemize}
\item[i)] the width for the processes (\ref{eq:charge}) and (\ref{eq:neutral})
is indeed reduced by orders of magnitude with respect to the free space and the
reduction grows with the mass number $A$, as expected;
\item[ii)] when the distorsion of the outgoing pion is accounted for by letting
the latter to move in an optical potential, then the reduction is somewhat
tempered, but of course it remains very substantial;
\item[iii)] finally of much relevance is that for $^{41}Ca$ the width for
decaying in the uncharged channel (\ref{eq:neutral}) becomes even larger than
the one for decaying in the charged channel (\ref{eq:charge}), thus possibly
indicating a violation of the $\Delta I=1/2$ rule.
\end{itemize}

\section{The $\Delta I=1/2$ rule}

A remarkable empirical finding in the $\Lambda$ decay relates to a selection
rule holding in isospace.
It is indeed found that the decay of the isospin zero ($I=0$) $\Lambda$ leads
overwhelmingly to a pion and a nucleon coupled to an isospin $I=1/2$ rather
than
to an $I=3/2$ state. The consequences, easy to work out, of this selection
rule on the decay width are

\be
\Gamma(\Lambda\rightarrow n\pi^0) = \frac{1}{2}
\Gamma(\Lambda\rightarrow p\pi^-)
\ee
or
\be
\frac{\Gamma(\Lambda\rightarrow charged \; hadrons)}
{\Gamma(\Lambda\rightarrow all \; hadrons)} = \frac{2}{3}
\ee
and are found to hold to within a few percent.
Yet violations occur and indeed one has just been quoted above for the mesonic
channel.
Further ones are expected to occur in the non-mesonic channel.

Furthermore, the theoretical foundation of the rule remains still now rather
obscure.

\section{Non-mesonic decay of the $\Lambda$: one nucleon induced process}

An alternative route for the decay of the $\Lambda$ in a nucleus was first
conjectured by Block and Dalitz as long ago as 1963 \cite{Blo63}.
It may be referred to as the {\em one nucleon induced decay}. Indeed it
corresponds to the occurrence of the process

\be
\Lambda + N \rightarrow N + N \;.
\ee
Assuming the energy to be equally shared by the two outgoing nucleons one has
for the energy of a nucleon

\be
E_N \simeq \frac{m_\Lambda - m_N}{2} \sim 89~MeV \;.
\ee
Hence

\be
p_N = \sqrt{2 m_N E_N} \simeq 400 \; MeV/c \gg k_F \;.
\ee

This channel of decay thus escapes the blocking induced by the Pauli principle
and indeed the experimental finding for the ratio of the widths associated with
the non-mesonic decay $\Gamma_{nm}$ versus the mesonic one $\Gamma_\pi$ yields

\be
\frac{\Gamma_{nm}}{\Gamma_\pi} \gg 1
\ee
already for $A\gapp 8\div 10$.

An interesting calculation of the non-mesonic one nucleon induced decay has
been
performed by Ramos and Bennhold \cite{Ram95} in the hypernucleus $^{12}_\Lambda
C$. These authors assume a weak meson-exchange interaction (namely having one
strong and one weak vertex) carried by $\pi, \eta, K, \rho, \omega$ and $K^*$
and also address the question of the nature of the partner nucleon assisting
the decay (namely whether it is a proton or a neutron). Their results
\cite{Ram95} deserve two comments:

\begin{itemize}
\item[i)] concerning the width it appears that, when all the above mentioned
mesons carrying the interaction are brought into the game, one obtains a value
essentially identical to the one obtained with the pion alone, which is however
still somewhat lower than the experimental value ($\Gamma_{nm}/\Gamma_{free}=0.
87$ to be compared with the experimental value $\Gamma_{exp}/\Gamma_{free}=1.
14\pm 0.20$);
\item[ii)] concerning the ratio $\Gamma_n/\Gamma_p$, where the width $\Gamma_n$
relates to the non-mesonic $\Lambda$ decay with a neutron partner whereas
$\Gamma_p$ has a proton partner, the theoretical prediction with the pion alone
is 0.19, whereas with the full interaction it grows to 0.3, still a long way
from the experimental value $1.33^{+1.12}_{-0.81}$.
\end{itemize}

Of significance is that only the lower bound of the experimental value appears
to overlap with the predictions of the $\Delta I=1/2$ rule, eventually
indicating a violation of the latter.

For a better appreciation of the related physics it is worth now to briefly
review how the $\Lambda$ decay width is actually calculated.
The basic formula is of course

\be
\Gamma_\Lambda = -2 Im~\Sigma_\Lambda \;,
\ee
$\Sigma_\Lambda$ being the $\Lambda$ self-energy.
Diagrammatically the process is illustrated in Fig.2, where the free and
the {\em in medium} decay are displayed.

In the free case clearly the free pion propagator

\be
{\cal D}_0(q,\omega) = \frac{1}{\omega^2-q^2-\omega_q^2} \; ,
\ee
where
\be
\omega_q = \sqrt{q^2+m_\pi^2} \; ,
\ee
should be used in calculating the $\Lambda$ self-energy.
One thus gets

\be
\Gamma_\Lambda = \frac{3}{\pi}\int_0^{\infty} dq
q^2\left(s_\pi^2+\frac{p_\pi^2}{m_\pi^2} q^2\right)
\frac{\delta(\omega-\omega_q)}{2\omega_q}
\label{eq:GL}
\ee
where both the $s$ and the $p$ waves of the emitted pion are permitted owing
to the parity non-conservation (actually the $s$ piece violates the parity, the
$p$ one conserves it).
To get the experimental value of the free $\Lambda$ decay one uses the values
\be
s_\pi=2.34\time 10^{-7} \;\;,\;\; p_\pi=1.17\times 10^{-7} \;.
\ee

\begin{figure}
\vspace{15cm}
\caption{ The $\Lambda$ self-energy $\Sigma$ in free space (a) and in the
medium (b). The content of the dashed bubble is largely dictated by the degree
of off-shellness of the pion.}
\end{figure}

The mesonic decay in the nucleus requires instead the $\pi$ propagator in the
medium, which reads

\be
{\cal D}(q,\omega) = \frac{1}{\omega^2-q^2-\omega_q^2-q^2\Pi(q,\omega)} \;,
\ee
$\Pi(q,\omega)$ being the so-called polarization propagator.

In order to describe the one nucleon induced $\Lambda$ decay, $\Pi(q,\omega)$
should embody the particle-hole (ph) excitations of the system.
For nuclear matter these cover the dashed region of Fig.3 and how they enter
into the non-mesonic decay channel is illustrated by the diagram displayed in
Fig.4a.

\begin{figure}
\vspace{15cm}
\caption{ The excitation spectrum of nuclear matter. Dashed region: the
p-h excitations. Also shown: the pionic branch in the vacuum and in the medium
($k_F=1.2 fm^{-1}$) and the $\Delta$-excitation.}
\end{figure}

It is clear that only a far off-shell pion can excite ph states
since the curve corresponding to the energy available for the emitted pion,
namely

\be
\omega = \Delta m -\frac{q^2}{2 m_N} \;
\label{eq:omega}
\ee
with $\Delta m = m_\Lambda - m_N$, overlaps with the allowed ph zone only for
pions far from the on-shell line (and also from the pionic branch in nuclear
matter).

Even when the difference in the potential energy felt by a $\Lambda$ and a
nucleon inside nuclear matter in taken into account in (\ref{eq:omega}),

\be
\omega = \omega_m = \Delta m + (V_\Lambda - V_N) - \frac{q^2}{2 m_N} \; ,
\label{eq:omm}
\ee
the above conclusion is not going to be much
changed.

Naturally pions emitted from the weak vertex can be far off-shell and thus the
one nucleon induced process opens an important door for the $\Lambda$ decay in
a
nucleus, as in fact the previously quoted calculation of Ramos and Bennhold
\cite{Ram95} indicates.

Additional support comes from the approach to the same problem
followed by Alberico {\em et al.} \cite{Alb91}.
These authors use the RPA expression for the polarization propagator, namely

\be
q^2\Pi_{ph}(q,\omega) = \frac{q^2\Pi^0(q,\omega)}{1-g'(q)\Pi^0(q,\omega)}
\ee
$\Pi^0(q,\omega)$ being the bare ph propagator, which however, unlike in ref.
\cite{Ram95}, also includes the $\Delta$ resonance, namely

\be
\Pi^0 = \Pi_0^N + \Pi_0^\Delta \; .
\label{eq:Pi0}
\ee
For the short range repulsion the Landau and Migdal type phenomenological
expression \cite{Ose82}

\be
g'(q) = \Gamma_\pi^2(q,\omega) \frac{1}{3} \frac{q^2+q_c^2}{q^2+q_c^2+m_\pi^2}
+ C_\rho \Gamma_\rho^2 \frac{2}{3} \frac{q^2+q_c^2}{q^2+q_c^2+m_\rho^2} \; ,
\label{eq:gprime}
\ee
with $q_c=600$ MeV, is utilized.

It is worth pointing out that $g'(q)$ couples the weak vertex not only to
pionic
modes, but also, via the parity conserving piece ($p$-wave), to the transverse
($\rho$-like) modes.
The above takes care of the strong vertex where the virtual pion is absorbed.

Concerning the weak vertex, where the virtual pion is emitted, Alberico {\em et
al.} \cite{Alb91} deduce the following expressions

\begin{eqnarray}
V^{(s)}(q,\omega) &=&
q^2 {\cal D}_0(q,\omega) -q^2 \tilde{\cal D}_0(q,\omega)
\\
V^{(p,L)} (q,\omega) &=& g'_\Lambda +
q^2 {\cal D}_0(q,\omega) - \frac{2}{3} q^2 \tilde{\cal D}_0(q,
\omega)
\\
V^{(p,T)} (q,\omega) &=& g'_\Lambda +
\frac{1}{3} q^2 \tilde{\cal D}_0(q,\omega)
\end{eqnarray}
where $g'_\Lambda$ is taken just as a parameter (positive) simply because at
present it cannot be computed and $\tilde{\cal D}_0(q,\omega)$ is the free pion
propagator with $q^2$ replaced by $q^2+q_c^2$.

The above vertices should then be inserted into the formula, easily derived
from
(\ref{eq:GL}),

\be
\Gamma^{(s)} = -\frac{3}{\pi^2} s_\pi^2 \int_{k_F}^{q_{max}} dq
\left[V^{(s)}(q,\omega)\right]^2 Im~\Pi_L(q,\omega) \mid_{\omega=\omega_m}
\; ,
\ee
to get the $s$-wave decay width, and into

\be
\Gamma^{(p,\sigma)} = -\frac{3}{\pi^2} \frac{p_\pi^2}{m_\pi^2}
\int_{k_F}^{q_{max}} dq
\left[V^{(p,\sigma)}(q,\omega)\right]^2 Im~\Pi_\sigma(q,\omega)
\mid_{\omega=\omega_m} \; ,
\ee
to get the $p$-wave one ($\omega_m$ is given by (\ref{eq:omm})).
In the above total Pauli blocking has been assumed (the lower limit of
integration is indeed $k_F$, the upper one being $q_{max}=\sqrt{2 m_N (\Delta m
+ V_\Lambda - V_M)}$, as it is straightforward to verify) and the index
$\sigma$ refers to both the longitudinal $(L)$ and transverse $(T)$ channel,
respectively.

The obtained results for the one-nucleon induced $\Lambda$ decay are displayed
in the first four row of Table 1 for a variety of choices for the parameter
$g'_\Lambda$, which of course affect only the parity-conserving $p$-channel.

\begin{center}
\begin{tabular}{|c|c|c|c|c|c|c|} \hline
 $g'_\Lambda$  & 0      & 0.1    & 0.2   & 0.3   & 0.4   &0.5   \\ \hline\hline
 $R_1^{(s)}$   & 0.36   & 0.36   & 0.36  & 0.36  & 0.36  &0.36  \\ \hline
 $R_1^{(p,L)}$ & 1.02   & 0.69   & 0.42  & 0.22  & 0.091 &0.024 \\ \hline
 $R_1^{(p,T)}$ & 0.008  & 0.012  & 0.096 & 0.26  & 0.51  &0.83  \\ \hline
 $R_1$         & 1.39   & 1.06   & 0.88  & 0.84  & 0.96  &1.22  \\ \hline
 $R_2^{(s)}$   & 0.40   & 0.40   & 0.40  & 0.40  & 0.40  &0.40  \\ \hline
 $R_2^{(p,L)}$ & 0.40   & 0.31   & 0.24  & 0.18  & 0.12  &0.079 \\ \hline
 $R_2^{(p,T)}$ & 0.0002 & 0.0009 & 0.005 & 0.013 & 0.025 &0.040 \\ \hline
 $R_2$         & 0.79   & 0.71   & 0.64  & 0.59  & 0.54  &0.52  \\ \hline
 $R=R_1+R_2$   & 2.18   & 1.77   & 1.52  & 1.43  & 1.50  &1.74  \\ \hline
\end{tabular}
\end{center}
Table 1. Various contributions to the ratio $R$ between the $\Lambda$ decay
widths in RPA-correlated nuclear matter and in the vacuum for different values
of $g'_A$ \cite{Alb91}. The second, third and fourth rows give the $s$-wave,
the
longitudinal and transverse $p$-wave and the total contribution respectively
for
the one-nucleon induced $\Lambda$ decay. The same quantities for the
two-nucleon induced $\Lambda$ decay are quoted in the next four rows.
In the final row the global ratio is displayed.

\vspace{1cm}
\noindent
Note that for $g'_\Lambda = 0.3 \div 0.4$ the quoted nuclear matter result is
essentially in accord with the previously reported one of ref.\cite{Ram95},
obtained with a sophisticated interaction in an hypernucleus as light as
$^{12}_\Lambda C$.
On the other hand Ramos, Oset and Salcedo \cite{Sal95} obtain for a variety of
nuclei (see Table 2) results substantially larger than those of Table 1, which
might be due to the fact that the calculations in ref.\cite{Sal95} is done in
finite nuclei in a local density approximation and with experimental single
particle energies (but ignoring the $\Delta$ contribution).

\vspace{1cm}
\begin{center}
\begin{tabular}{|c|c|c|c|} \hline
  & $\Gamma_\pi$ & $\Gamma_{1p1h}$ & $\Gamma_{2p2h}$
 \\ \hline\hline
 $^{12}_\Lambda C$   & 0.31 & 1.45 & 0.27 \\ \hline
 $^{16}_\Lambda O$   & 0.24 & 1.54 & 0.29\\ \hline
 $^{20}_\Lambda Ne$  & 0.14 & 1.60 & 0.32\\ \hline
 $^{40}_\Lambda Ca$  & 0.03 & 1.76 & 0.32\\ \hline
 $^{56}_\Lambda Fe$  & 0.01 & 1.82 & 0.32\\ \hline
 $^{89}_\Lambda Y$   &  -   & 1.88 & 0.31\\ \hline
 $^{100}_\Lambda Ru$ &  -   & 1.89 & 0.31\\ \hline
 $^{208}_\Lambda Pb$ &  -   & 1.93 & 0.30\\ \hline
\end{tabular}
\end{center}
Table 2. Hypernuclear decay widths according to \cite{Sal95}.
$\Gamma_\pi$ is the mesonic width, $\Gamma_{1p1h}$ is the width for the decay
(one-nucleon induced) into a two-nucleon final state whereas $\Gamma_{2p2h}$
(two-nucleon induced) relates to a final state with three nucleons.
\vspace{1cm}

\section{Non-mesonic decay of the $\Lambda$: two nucleons induced process}

Alberico {\em et al.} \cite{Alb91} conjectured in 1990 that an alternative
non-mesonic decay channel for the $\Lambda$ in a nucleus could be offered by
letting the pion to be absorbed by a correlated pair of nucleons (or to excite
a
2p-2h state), as illustrated in Fig.4b.

\begin{figure}
\vspace{15cm}
\caption{ Schematic representation of the $\Lambda$ decay: in the case a) the
decay is driven by the coupling with a p-h excitation; in the case b) the decay
is driven by the coupling with a 2p-2h excitation.
}
\end{figure}

This process goes in parallel to the related absorption of a real photon of,
say, 100 MeV of energy, whose occurrence on a single nucleon in a nucleus can
hardly meet the requirements of energy and momentum conservation.

A glance at Fig.3 is sufficient for concluding that pions emitted close to the
mass shell preferentially select this channel to be absorbed, thus allowing the
$\Lambda$ to decay.

One has to realize that, in the medium, the pion dispersion relation
corresponding to the pionic branch of Fig.3 would intercept (\ref{eq:omm})
(with $V_\Lambda - V_N =20~MeV$) still in the Pauli blocked region, namely at
$q\simeq 1.09~fm^{-1}$. However the pionic branch in the medium has a width
associated precisely with the 2p-2h excitations above referred to and, owing to
this width, the decaying $\Lambda$ can escape the constraints arising from the
Pauli principle.

Indeed in the energy domain $\omega\simeq m_\pi$ the pionic branch decays via
2p-2h emission and thus overlaps with the parabola (\ref{eq:omm}) also {\em
outside the Pauli blocked domain}.
Obviously the $\Lambda$ decay in this instance results in a 3p-2h final state,
as shown in Fig.4b.

The pioneering calculation of this process carried out in ref.\cite{Alb91}
assumed

\be
Im~\Pi^{*}_{2p-2h} \ne 0 \;\;\; \mbox{only when} \;\; Im~\Pi_{ph}=0 \; ,
\ee
{\em i.e.} the complicated physics associated with the interference between
2p-2h and 1p-1h excitations is ignored: accordingly the 2p-2h
contribution to the $\Lambda$ decay in nuclear matter was restricted
to the range

\be
k_F \leq q \leq 1.65~fm^{-1} \; .
\ee
In addition the following parametrization
\be
\Pi^{*}_{2p-2h} = - 4\pi\rho^2 {\cal C}^0
\ee
for the irreducible part of the polarization propagator was adopted,
$\rho=2k_F^3/3\pi^2$ being the nuclear matter density.
The experimental information from quasi-elastic electron scattering $(e,e')$
and from the mesic atoms suggests the value
\be
Im~{\cal C}^0 \simeq 0.18~m_\pi^{-6}
\ee
for the imaginary part and
\be
Re~{\cal C}^0 \simeq 0.40~m_\pi^{-6}
\ee
for the real part.

Utilizing these inputs one then writes for the polarization propagator

\be
q^2\Pi(q,\omega) = \frac{q^2\left[\Pi^0(q,\omega)+4\pi\rho^2 {\cal C}^0\right]}
{1-g'(q)\left[\Pi^0(q,\omega)+4\pi\rho^2{\cal C}^0\right]}
\label{eq:Pi}
\ee
with $\Pi^0(q,\omega)$ and $g'(q)$ given by (\ref{eq:Pi0}) and
(\ref{eq:gprime}).

The results obtained with this approach, admittedly crude, are quoted in Table
1
and indeed point to a sizable width for the $\Lambda$ decay associated with
$2p-2h$ excitations.

Ramos, Oset and Salcedo \cite{Sal95} revisited the subject with a more refined
calculation.
In particular they took

\be
\Pi^{*}_{2p-2h}(q,\omega) = -4\pi q^2 {\cal C}^{*}_0 \rho^2
\ee
with
\be
{\cal C}^{*}_0\rho^2 = (0.105 + i 0.096)~m_\pi^{-6} \; ,
\ee
whose imaginary and real part are about half and a quarter, respectively, of
(\ref{eq:Pi}).

In addition the calculation of Ramos {\em et al.} has been performed in
finite nuclei via the local density approximation, but again ignoring the
$\Delta$ excitation.
Yet their results, while of course smaller, are not much
at odd with the one of Table 1 and confirm, anyway, the importance of this
channel for the decay of the $\Lambda$ in a nucleus.

Of particular significance are the implications of the two-nucleon induced
$\Lambda$ decay for the ratio $\Gamma_n/\Gamma_p$, a subject which is currently
much explored and likely to provide interesting clues on the validity of the
$\Delta I=1/2$ rule.

\end{document}